\documentclass[lettersize,journal]{IEEEtran}
\usepackage{amsmath,amsfonts}
\usepackage{algorithmic}
\usepackage{algorithm}
\usepackage{array}
\usepackage[caption=false,font=normalsize,labelfont=sf,textfont=sf]{subfig}
\usepackage{textcomp}
\usepackage{stfloats}
\usepackage{url}
\usepackage{verbatim}
\usepackage{graphicx}
\usepackage{cite}

\usepackage{amsmath,graphicx}
\usepackage{cite}
\usepackage{amssymb,amsfonts,amsthm,bm}
\usepackage[usenames,dvipsnames]{xcolor}

\usepackage{booktabs}
\usepackage{graphicx}
\usepackage{multirow}
\usepackage{stfloats}
\usepackage{array} 
\usepackage{caption}

\begin{document}

\title{A Two-Stage Deep Representation Learning-Based Speech Enhancement Method Using Variational Autoencoder and Adversarial Training}

\author{Yang~Xiang,~\IEEEmembership{Student Member,~IEEE,}
        Jesper~Lisby~Højvang,
        Morten~Højfeldt~Rasmussen,
        and~Mads~Græsbøll~Christensen,~\IEEEmembership{Senior Member,~IEEE}


\thanks{Y. Xiang is an industrial Ph.D. student, associated with the Audio Analysis Lab, CREATE, Aalborg University, Aalborg, Denmark, and Capturi A/S, Aarhus, Denmark; email: yaxi@create.aau.dk}
\thanks{J. L. Højvang and M. H. Rasmussen are with Capturi A/S, Aarhus, Denmark; email: \{jlh,mhr\}@capturi.com} 
\thanks{M. G. Christensen is with the Audio Analysis Lab, CREATE, Aalborg University, Aalborg, Denmark; email: mgc@create.aau.dk}
\thanks{This work was supported by Innovation Fund Denmark (Grant No. 9065-00046).}
}

\markboth{Journal of \LaTeX\ Class Files,~Vol.~14, No.~8, March~2022}%
{Shell \MakeLowercase{\textit{et al.}}: A Sample Article Using IEEEtran.cls for IEEE Journals}
\maketitle

\begin{abstract}

This paper focuses on leveraging deep representation learning (DRL) for speech enhancement (SE). In general, the performance of the deep neural network (DNN) is heavily dependent on the learning of data representation. However, the DRL's importance is often ignored in many DNN-based SE algorithms. To obtain a higher quality enhanced speech, we propose a two-stage DRL-based SE method through adversarial training. In the first stage, we disentangle different latent variables because disentangled representations can help DNN generate a better enhanced speech. Specifically, we use the $\beta$-variational autoencoder (VAE) algorithm to obtain the speech and noise posterior estimations and related representations from the observed signal.  However, since the posteriors and representations are intractable and we can only apply a conditional assumption to estimate them, it is difficult to ensure that these estimations are always pretty accurate, which may potentially degrade the final accuracy of the signal estimation. To further improve the quality of enhanced speech, in the second stage, we introduce adversarial training to reduce the effect of the inaccurate posterior towards signal reconstruction and improve the signal estimation accuracy, making our algorithm more robust for the potentially inaccurate posterior estimations. As a result, better SE performance can be achieved. The experimental results indicate that the proposed strategy can help similar DNN-based SE algorithms achieve higher short-time objective intelligibility (STOI), perceptual evaluation of speech quality (PESQ), and scale-invariant signal-to-distortion ratio (SI-SDR) scores. Moreover, the proposed algorithm can also outperform recent competitive SE algorithms.



\end{abstract}


\section{Introduction}
\IEEEPARstart{I}{n} real-world environments, speech signals are usually degraded by various environmental noise. To counter these degradations, speech enhancement (SE) techniques have been developed during the past decades \cite{loizou2013speech}. The main purpose of SE is to remove background noise from an observed signal and improve speech quality and intelligibility in a noisy environment. SE has been widely applied in speech coding, teleconferencing, hearing aids, mobile communication, and robust automatic speech recognition (ASR) \cite{li2014overview}. Due to the recent COVID-19 pandemic, there has been an increasing need for online meeting systems \cite{pandey2020impact}, where SE can help the system to reduce the word error rate (WER) for accurate live captioning when transmitting high-quality speech audio in various complex-noise conditions \cite{eskimez2021human,iwamoto2022bad}. Therefore, SE is an increasingly prominent research topic.

There is a considerable amount of literature published on SE algorithms. Classic SE methods include signal subspace methods \cite{jabloun2003incorporating,jensen2015noise,christensen2016experimental}, codebook-based methods \cite{srinivasan2005codebook,kavalekalam2018model,kavalekalam2018online}, and non-negative matrix factorization (NMF) methods \cite{mohammadiha2013supervised,kavalekalam2018online,xiang2020nmf,xiang2021novel}. Most of these methods perform SE by applying short-time Fourier transform (STFT) to analyze the time--frequency (T--F) representation of the observed signal or directly using waveform. Recently, with the development of deep neural network (DNN) techniques, DNNs have shown a great potential for SE \cite{xu2014regression,wang2014training, wang2018supervised,sun2017multiple,luo2019conv, xiang2020parallel,hu20g_interspeech, li2021icassp, wang2021compensation}. {These DNN-based SE methods usually apply different structures (e.g. feedforward multilayer perceptron (MLP) \cite{xu2013experimental,xu2014regression}, convolutional neural network (FCN) \cite{park2016fully}, and deep recurrent neural networks {(DRNN)} \cite{jacobsson2005rule,huang2015joint, keshavarzi2018use, goehring2019using}) to predict various targets \cite{wang2018supervised}.} Unlike classic algorithms \cite{jensen2015noise, kavalekalam2018model,xiang2021novel,mohammadiha2013supervised,kavalekalam2018online,xiang2020nmf}, DNNs can learn the disentangled representations of the data \cite{chan2105white}, and can use the learned representations to generate related data. Thus, we hypothesize that one of the reasons of why DNN can perform SE is that DNN can extract useful speech representation~\cite{wright2022high} from the observed signal and generate corresponding speech data. DNNs' advantage for SE is that DNN can extract underlying information (e.g., phoneme or emotional information) from high dimension features~\cite{hsu2017learning, hsu2017unsupervised, hsu2017unsupervised_learning, xie2021disentangled}. Moreover, DNN can also represent the different underlying information by different vector forms, and can disentangle different information.  As a result, DNNs can effectively analyze more signal representations and achieve a better SE performance. Additionally, one of the DNNs' principles is that DNNs are based on data representation learning~\cite{chan2105white,bengio2013representation, dai2021closed}, so it can avoid the speech-phase estimation problem (only DNN's input contains the all signal information) \cite{williamson2015complex,gerkmann2015phase, fu2017raw} in traditional T--F processes (STFT analysis). More specifically, recent research \cite{fu2017raw} has indicated that DNN can directly leverage the speech waveform to achieve excellent SE performance \cite{fu2018end}. 
Furthermore, compared to T--F representations, DNNs can easily combine different information to perform the signal analysis (find underlying relationships of different signals), so the audio--visual-based SE has also been developed in recent years~\cite{michelsanti2021overview, carbajal2021disentanglement}.

Currently, although DNNs have significantly promoted the development of SE techniques \cite{wang2018supervised}, there are still some problems in DNN-based SE algorithms. The DNNs' potential for SE is not completely explored. For example, most of the present DNN-based SE methods \cite{xu2013experimental,xu2014regression,wang2014training, wang2018supervised,luo2019conv, xiang2020parallel,hu20g_interspeech, li2021icassp, wang2021compensation} focus on the learning of the training target and apply DNNs only to predict pre-defined targets (e.g., various masks \cite{wang2014training}, speech spectrum \cite{xu2013experimental}, and speech present probability \cite{tu2019speech}). However, these algorithms ignore the importance of reliable representations for DNN-based methods \cite{bengio2013representation} and do not consider using DNN to obtain better signal representations. Although direct prediction of pre-defined targets can prevent inaccurate signal assumptions \cite{xu2013experimental}, the lack of a good representation learning strategy means that these algorithms do not achieve constant satisfactory SE performance in complex noisy environments \cite{wang2018supervised}. On the contrary, an efficient deep representation learning (DRL) method may not only improve DNNs' ability to extract useful information in complex environments~\cite{xie2021disentangled,bengio2013representation} but can also lead to a better prediction ability of the DNN  \cite{bengio2013representation}. Moreover, a good representation can place less demand on the learning machine in order to perform a task successfully \cite{wang2018supervised}. Therefore, DRL has potential to help DNN-based SE algorithms improve their robustness and generalization ability \cite{bengio2013representation, wright2022high}. Furthermore, DRL can disentangle different latent representations of the speech signal (e.g., content and acoustic representation) \cite{hsu2017learning, hsu2017unsupervised, hsu2017unsupervised_learning}, so more related information (e.g., phonetic information of a speech signal) can be included to analyze the speech signal when performing SE, which has a significant potential to improve the quality and intelligibility of the enhanced speech.{
DRL plays a crucial role in finding, disentangling, and analyzing intricate speech information during SE, thereby endowing DRL-based SE algorithms with the potential to reduce WER in ASR systems while improving the human listening experience. This potential stems from the ability of DRL-based methods to analyze various speech-related information and mitigate information loss caused by speech distortion, a capability that previous SE algorithms did not possess \cite{wang2018supervised}. As a result, the DRL-based SE strategy holds promise for applications in hearing aids, robust ASR systems, and online meeting systems where reducing WER and achieving accurate live captioning is crucial when transmitting high-quality speech audio.}

Due to the importance of DRL for DNN \cite{bengio2013representation, dai2021closed}, DRL-based SE algorithms have been investigated in recent research works \cite{leglaive2018variance,bando2018statistical,leglaive2019semi,carbajal2021guided,fang2021variational, carbajal2021disentanglement}. These methods mainly use a variational autoencoder (VAE) \cite{kingma2013auto} to learn speech representations and improve the generalization ability of the algorithms. VAE is a DRL model that can make efficient approximate posterior inferences and learn the probability distribution of complex data. Therefore, VAE can help DNN extract useful information from the signals \cite{kingma2013auto}. Currently, VAE has been widely applied in various tasks related to representation learning \cite{kim2021conditional,zhang2021visinger}. Although such VAE-based SE algorithms effectively improve DNN's generalization ability, they only consider the speech representation of the observed signal and do not attempt to disentangle the speech representation with latent noise representations.  Instead, they use NMF to model the noise signal \cite{leglaive2018variance,bando2018statistical,leglaive2019semi,carbajal2021guided,fang2021variational, carbajal2021disentanglement}. This directly results in inaccurate obtained speech representations and possibly unsatisfactory SE performance \cite{bando2018statistical}.

To obtain a more accurate speech representation, a novel VAE-based SE method  \cite{xiang2022bayesian}, named  Bayesian permutation training variational
autoencoder (PVAE), was proposed in our preliminary research. This method leverages a conditional posterior assumption to derive a novel evidence lower bound (ELBO) that enables the VAE to disentangle different signal representations in a very effective way. In addition, the derived ELBO also leads to a novel VAE model for SE. Compared to previous VAE-based SE models \cite{leglaive2018variance,bando2018statistical,leglaive2019semi,carbajal2021guided,fang2021variational, carbajal2021disentanglement}, this model first extracts a more accurate speech representation from the observed signal, because different latent representations are disentangled \cite{xiang2022bayesian} and these representations are expressed in a low-dimension space; the extracted representations are then used as the input of different decoders for SE. PVAE \cite{xiang2022bayesian} can be directly adopted by many current SE DNN structures \cite{wang2018supervised} and also directly used to optimize DNN-based SE algorithms \cite{wang2018supervised}. Conducted experiments \cite{xiang2022bayesian} indicate that this DRL strategy can help the traditional DNN-based SE method \cite{huang2014deep} achieve a better SE performance.

To further help PVAE to achieve better SE performance, we propose to leverage $\beta$-VAE \cite{higgins2016beta, burgess2018understanding} to improve PVAE's  representation learning ability. More specifically, the proposed $\beta$-PVAE \cite{xiang2022deep} algorithm improves PVAE's capacity to disentangle different latent variables from the observed signal, which means that $\beta$-PVAE can obtain a better signal representation for SE. Moreover, $\beta$-PVAE optimizes the PVAE's network structure by setting $\beta$ to infinity, which ensures that $\beta$-PVAE can not only improve PVAE's SE performance but also reduce the number of PVAE training parameters.

Both the speech and noise signal representations obtained by PVAE and $\beta$-PVAE are based on speech and noise posterior estimations \cite{xiang2022bayesian}. An experimental analysis in \cite{xiang2022deep} indicated that an accurate posterior estimation is 
crucial for $\beta$-PVAE because $\beta$-PVAE's decoders rely heavily on the accurate representation as input to reconstruct signals. Therefore, an accurate posterior estimation can lead to high SE performance \cite{xiang2022deep}. On the other hand, an inaccurate posterior estimate can undermine the decoder's SE performance. However, obtaining pretty accurate posterior estimations is difficult since posteriors are intractable. In addition, another possible reason for the potential inaccurate posterior estimation is that the posterior estimations in \cite{xiang2022bayesian} rely on a conditional assumption \cite{xiang2022deep}. Although this conditional assumption results in a good signal model and ensures that various signal representations can be disentangled, it is difficult to validate that this assumption is consistently correct in a complex noisy environment. As a result, it potentially leads to  $\beta$-PVAE to have inaccurate  speech signal estimate and its SE performance is limited. 

To mitigate the effect of inaccurate posterior estimations for the signal estimation and improve the SE performance of our preliminary work \cite{xiang2022deep}, we extend our DRL-based SE framework \cite{xiang2022bayesian,xiang2022deep} and propose in this paper a two-stage DRL-based SE method consisting of a representation learning stage \cite{bengio2013representation} and an adversarial training stage \cite{goodfellow2014generative}. In the first representation learning stage, we leverage the $\beta$-PVAE \cite{xiang2022deep} to disentangle different signal representations from the observed signal to obtain speech and noise representations from the observed signal. To further obtain a better SE performance, in the second adversarial training stage, we propose to leverage generative adversarial networks (GANs) to improve the decoders' robustness for any possible inaccurate posterior estimation. Because we cannot ensure that the obtained posterior estimations are always accurate using $\beta$-PVAE, we instead attempt to make the decoders more robust. GAN is a probability generative model which can perform exact sampling from the desired distribution given random variables as input, using different $f$-divergence as training metrics  \cite{goodfellow2014generative,nowozin2016f}. Unlike the $\beta$-PVAE's decoder, this model can generate a desired sample without having precise knowledge of the distribution of the input sample. Moreover, adversarial training can usually improve VAE decoder's signal reconstruction ability and help the VAE obtain higher quality signals \cite{larsen2016autoencoding, parmar2021dual, huang2018introvae, kim2021conditional, zhang2021visinger}. Therefore, we introduce adversarial training to improve $\beta$-PVAE decoders' generative ability.


Recently, a combination of VAE and GAN (VAE--GAN) \cite{larsen2016autoencoding, parmar2021dual, huang2018introvae} has been widely applied in various speech synthesis tasks \cite{kim2021conditional, zhang2021visinger}. VAE--GAN can achieve better performance than independent GAN or VAE-based methods \cite{larsen2016autoencoding}, which usually use VAE to obtain a reliable signal representation and then involve the GAN to generate a high-quality signal. However, unlike our VAE--GAN-based SE algorithm, most of the current VAE--GAN-based methods \cite{larsen2016autoencoding, kim2021conditional, zhang2021visinger} do not disentangle various representations in the VAE training stage. To the best of our knowledge, this is the first attempt to investigate VAE--GAN's application in the SE field. Furthermore, compared to the current competitive GAN-based SE methods \cite{pascual2017segan, michelsanti2017conditional}, VAE--GAN can obtain a disentangled signal representation as the GAN's input. A discriminative input can place less demand on the learning machine in order to perform a task successfully \cite{wang2018supervised}, which means that our VAE-GAN can help current GAN-based SE algorithms generate a higher quality speech signal. 

This paper is organized as follows. First, in Section II, we will briefly review related VAE and GAN works. Then, we will proceed to illustrate the proposed two-stage VAE--GAN-based SE method in Section III and the experimental preparation, comparison, and analysis in Section IV. Finally, we draw conclusions in Section V.

\section{Fundamentals}

\subsection{Signal Model}

\begin{figure}[!tbp]
  \centering
  \centerline{\includegraphics[scale=0.7]{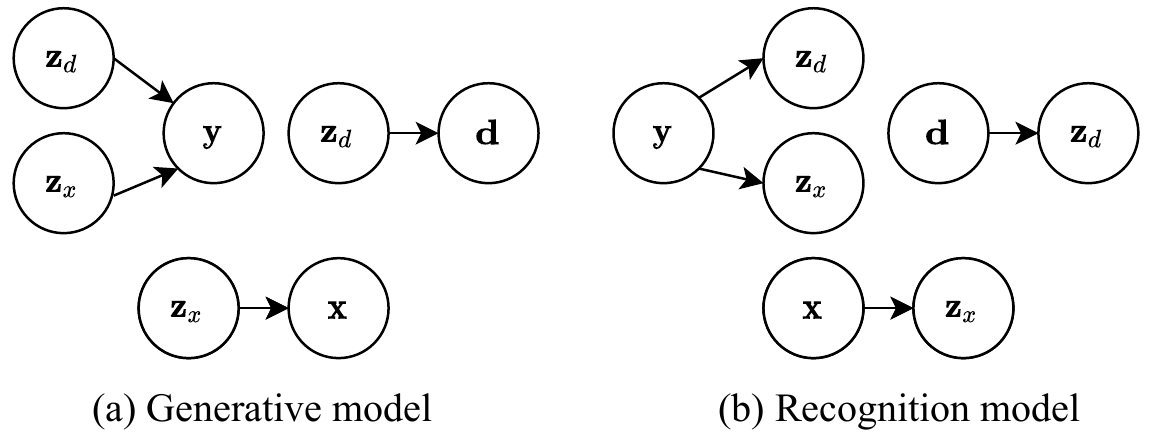}}
  \caption{Graphic illustration of the proposed signal model {\cite{xiang2022bayesian, xiang2022deep}}.}
  \label{fig:Bayesian_model}
\end{figure}

In this work, we assume that the noisy speech is additive, so the signal model can be written as follows:
\begin{equation}
  y(t) = x(t)+d(t),
  \label{time_noisy_model}
\end{equation}
where \(y(t)\), \(x(t)\), and \(d(t)\) represent the observed,  speech, and noise signal, respectively, and $t$ is the time index. Using the STFT, the observed signal $y_{f,n} \in \mathbb{C}$, speech signal $x_{f,n} \in \mathbb{C}$, and noise $d_{f,n} \in \mathbb{C}$ can be represented as 
\begin{equation}
  y_{f,n} = x_{f,n}+d_{f,n},
  \label{tf_noisy_model}
\end{equation}
where time frame index $n \in [1, N]$, and the frequency bin $f \in [1, F]$. $N$ and $F$ are the number of time frames and frequency bins, respectively. 

We use the log-power spectrum (LPS) as the DNN's input feature since LPS is thought to offer perceptually relevant parameters for DNN-based SE algorithms \cite{wan1999networks,xie1994family,xu2014regression, wang2018supervised}. At present, LPS, as the input feature, has been widely applied in the DNN-based SE algorithms \cite{wan1999networks,xie1994family,xu2014regression, wang2018supervised}. The LPS vector \cite{xu2014regression} at each frame is written as $\mathbf{y}$, $\mathbf{x}$, and $\mathbf{d}$, respectively (we omit the frequency and time frame index for simplicity). Moreover, in the following derivations of our algorithm, the additive assumption in models (\ref{time_noisy_model}) and (\ref{tf_noisy_model}) are not used. The purpose of (\ref{time_noisy_model}) and (\ref{tf_noisy_model}) is used to generate noisy signal. Furthermore, (\ref{time_noisy_model}) is a simple noisy signal model, so it is convenient to verify the correctness of our methods. Our framework has potential to analyze more challenging noisy signal models.

{Following our preliminary work \cite{xiang2022bayesian, xiang2022deep},} we assume that $\mathbf{y}$ can be generated from a random process involving the speech latent variables $\mathbf{z}_x \in {\mathbb{R}}^L$ and the noise latent variables $\mathbf{z}_d \in {\mathbb{R}}^L$ ($L$ is the dimension of latent variables). The latent variables $\mathbf{z}_x$ and $\mathbf{z}_d$ are independent representations of the speech and noise signal, respectively. The combination of $\mathbf{z}_x$ and $\mathbf{z}_d$ is the representation of the observed signal \cite{kingma2013auto,bengio2013representation}. The $\mathbf{x}$ and $\mathbf{d}$ can be independently generated by $\mathbf{z}_x$ and $\mathbf{z}_d$, respectively: the generative process is shown in Fig.~\ref{fig:Bayesian_model}(a). To obtain the latent variables $\mathbf{z}_x$ and $\mathbf{z}_d$, we assume that $\mathbf{z}_x$ and $\mathbf{z}_d$ can be estimated from the speech and noise posterior distributions $p(\mathbf{z}_x|\mathbf{x})$ and $p(\mathbf{z}_d|\mathbf{d})$, respectively, or from the noisy speech posterior distributions $p(\mathbf{z}_x|\mathbf{y} )$ and $p(\mathbf{z}_d|\mathbf{y})$ \cite{xiang2022bayesian}, based on the VAE's property \cite{kingma2013auto}. Fig.~\ref{fig:Bayesian_model}(b) shows the recognition process \cite{kingma2013auto}. To perform SE, it is necessary to disentangle the different latent variables from the observed signal. To simplify the disentanglement, we assume that $p({\mathbf{z}_x},{\mathbf{z}_d}|\mathbf{y})=p(\mathbf{z}_x|\mathbf{y})p(\mathbf{z}_d|\mathbf{y})$ in \cite{xiang2022bayesian}.


\subsection{VAE and $\beta$-VAE}

The original VAE is a probabilistic generative model \cite{kingma2013auto} which defines a probabilistic generative process between the observed signal and its latent variables and provides a principled method to jointly learn latent variables and generative and recognition models. Generative and recognition models are jointly trained by maximizing the ELBO  or minimizing the Kullback–Leibler (KL) divergence between their real joint distribution and the corresponding estimation \cite{kingma2013auto} using the stochastic gradient descent (SGD) or Adagrad~\cite{duchi2011adaptive} algorithm. Maximized, the ELBO can be written as follows: \begin{equation}
 \begin{aligned}
   &  {\mathbb E_{{\bf {y}} \sim p(\bf {y})}[\log q(\bf {y})]} \ge -{\mathcal{L}}_n, \\
   & {\mathcal{L}}_n = {\mathbb E_{{\bf {y}} \sim p(\bf {y})}} \left[D_{KL}\left({p({\bf z}_y|{\bf{y}}))}||{q({{\bf z}_y})}\right)\right] \\
   & \quad \quad - {\mathbb E_{{\bf {y}} \sim p(\bf {y})}} \left[ {\mathbb E_{{{\bf {z}}_y} \sim p({{\bf {z}}_y}|\bf {y})}}\left[\log {q({\bf{y}}|{\bf z}_y)} \right]\right],
   \end{aligned}
  \label{ELBO}
\end{equation}
where $D_{KL}(||)$ denotes the KL divergence; $\mathbf{z}_y \in {\mathbb{R}}^L$ is the noisy latent variable. Maximizing this lower bound is equivalent to minimizing ${\mathcal{L}}_n$.

Furthermore, $\beta$-VAE \cite{higgins2016beta,burgess2018understanding} is a modification of the original VAE framework, which introduces an adjustable hyperparameter $\beta$ in the KL divergence term: 
\begin{equation}
 \begin{aligned}
   & {\mathcal{L}}_{\beta} = \beta{\mathbb E_{{\bf {y}} \sim p(\bf {y})}} \left[D_{KL}\left({p({\bf z}_y|{\bf{y}}))}||{q({{\bf z}_y})}\right)\right] \\
   & \quad \quad - {\mathbb E_{{\bf {y}} \sim p(\bf {y})}} \left[ {\mathbb E_{{{\bf {z}}_y} \sim p({{\bf {z}}_y}|\bf {y})}}\left[\log {q({\bf{y}}|{\bf z}_y)} \right]\right].
   \end{aligned}
  \label{beta_ELBO}
\end{equation}
$\beta$-VAE aims to help the original VAE \cite{kingma2013auto} to obtain a better signal representation. In general, $\beta > 1$ results in more disentangled latent representations \cite{higgins2016beta}. A higher value of $\beta$ can encourage learning a more disentangled representation. 


\subsection{PVAE}
\begin{figure*}[!tbp]
  \centering
  \centerline{\includegraphics[scale=0.75]{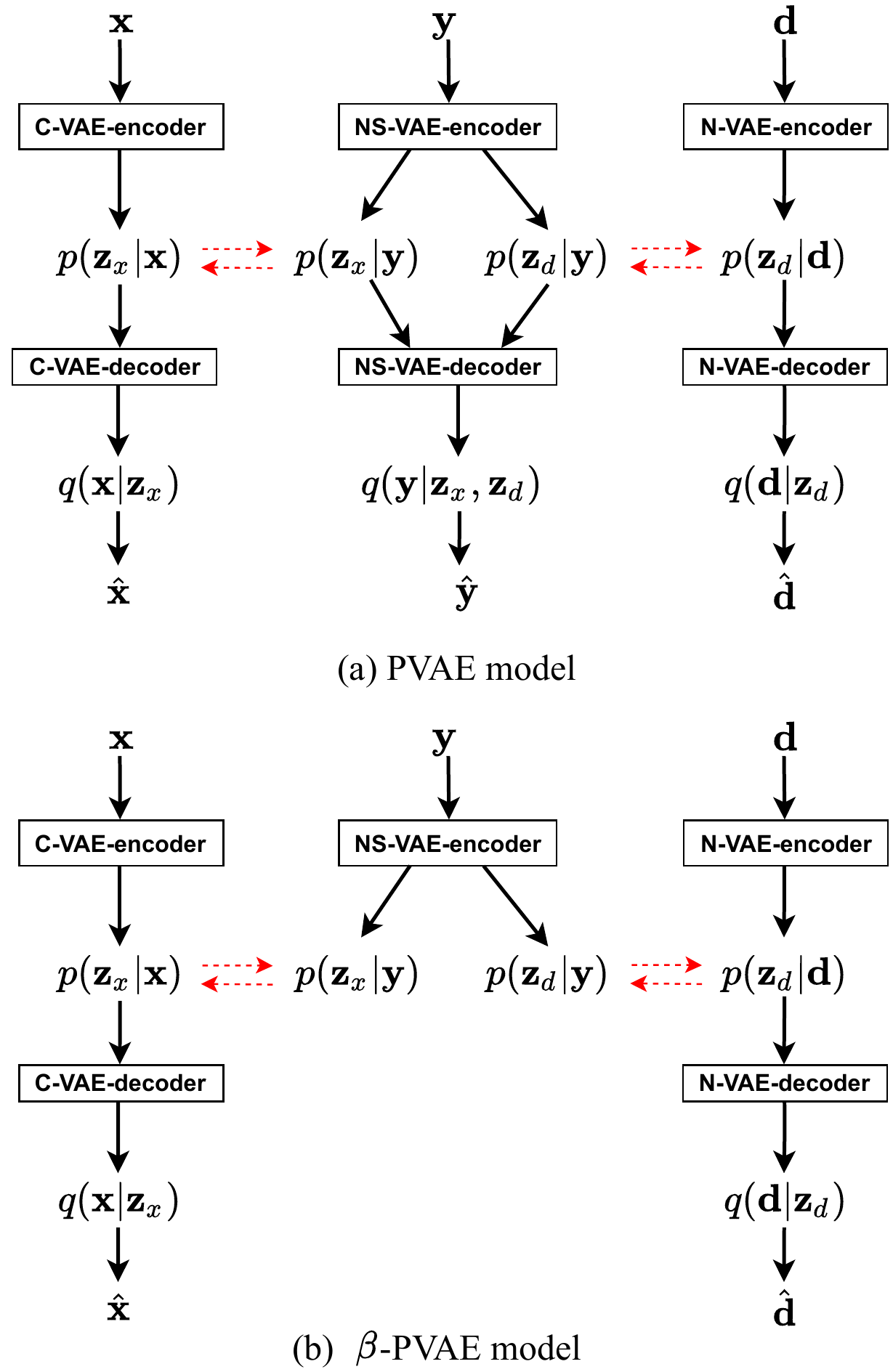}}
  \caption{Model illustration of PVAE and $\beta$-PVAE {\cite{xiang2022bayesian, xiang2022deep}}.}
  \label{fig:Bayesian_DNN}
\end{figure*}

Our preliminary work proposed a PVAE-based SE algorithm \cite{xiang2022bayesian} and indicated that PVAE can help the current DNN-based SE method \cite{huang2014deep} obtain better signal representations (because different latent representations are disentangled \cite{xiang2022bayesian} and these representations are expressed in a low-dimension space \cite{bengio2013representation}) and achieve better SE performance. PVAE is a semi-supervised DRL-based SE method which introduces multiple latent variables in VAE and disentangles them in a semi-supervised way for SE application. Fig.~\ref{fig:Bayesian_DNN}(a) shows the PVAE framework {\cite{xiang2022bayesian}}. We can see that PVAE includes three VAE structures: clean speech VAE (C-VAE), noise VAE (N-VAE), and noisy VAE (NS-VAE).  C-VAE and N-VAE are trained to obtain speech and noise latent representations and their posterior estimates $p({\bf z}_x|{\bf{x}})$, and $p({\bf z}_d|{\bf{d}})$, respectively. This is achieved by minimizing the following VAE loss function \cite{kingma2013auto}: 
\begin{equation}
 \begin{aligned}
   \mathcal{L}_{c} (\theta_x, \varphi_x; {\bf x}) &= {\mathbb E_{{\bf {x}} \sim p({\bf {x}})}} \{ D_{KL}\left({p({{\bf {z}}_x}|{\bf{x}})}||{q({\bf z}_x)}\right) \\
  & \quad - {\mathbb E_{{\bf {z}}_x \sim p({{\bf {z}}_x}|{\bf{x}})}} [\log {q({\bf x}|{\bf z}_x)} ]\},
   \end{aligned}
  \label{clean_vae}
\end{equation}

\begin{equation}
 \begin{aligned}
   \mathcal{L}_{d} (\theta_d, \varphi_d; {\bf d}) &= {\mathbb E_{{\bf {d}} \sim p({\bf {d}})}} \{ D_{KL}\left({p({{\bf {z}}_d}|{\bf{d}})}||{q({\bf z}_d)}\right) \\
  & \quad - {\mathbb E_{{\bf {z}}_d \sim p({{\bf {z}}_d}|{\bf{d}})}} [\log {q({\bf d}|{\bf z}_d)} ]\},
   \end{aligned}
  \label{noise_vae}
\end{equation}
where $\theta_x, \varphi_x, \theta_d, \varphi_d$ {are} the DNN parameters for the related probability estimation \cite{xiang2022bayesian}: $\theta_x$ and $\varphi_x$ are the C-VAE's encoder and decoder parameters, respectively; $\theta_d$ and $\varphi_d$ are the N-VAE's encoder and decoder parameters, respectively. {In this paper, we assign the symbol $\theta$ to represent the encoder-related parameters, while the symbol $\varphi$ is used to represent the decoder-related parameters.}  NS-VAE is trained under the supervision of C-VAE and N-VAE's encoders and is meant to disentangle speech and noise latent variables from the observed signal for SE application. Based on the derivation in \cite{xiang2022bayesian}, the NS-VAE's training loss function is expressed as follows:
\begin{equation}
 \begin{aligned}
  & \mathcal{L}_{p} (\theta_y, \varphi_y; {\bf y}) \\
  & \quad = {\mathbb E_{{\bf {y}} \sim p(\bf {y}),{\bf {x}} \sim p(\bf {x})}} \{D_{KL}\left({p({\bf z}_x|{\bf{y}})}||{p({\bf z}_x|{\bf{x}})}\right) \\
  & \quad \quad + {\mathbb E_{{{\bf {z}}_x} \sim p({{\bf {z}}_x}|{\bf {y}})}}[\log \frac{p({\bf z}_x|{\bf x})}{q({\bf z}_x)}]\} \\
  & \quad \quad + {\mathbb E_{{\bf {y}} \sim p({\bf {y}}), {\bf {d}} \sim p({\bf {d}})}} \{D_{KL}\left({p({\bf z}_d|{\bf{y}})}||{p({\bf z}_d|{\bf{d}})}\right) \\
  &  \quad \quad + {\mathbb E_{{{\bf {z}}_d} \sim p({{\bf {z}}_d}|{\bf {y}})}}[\log \frac{p({\bf z}_d|{\bf d})}{q({\bf z}_d)}]\} \\
  & \quad \quad - {\mathbb E_{{\bf {y}} \sim p(\bf {y})}} \left[ {\mathbb E_{{{\bf {z}}_d,{\bf {z}}_x} \sim p({{\bf {z}}_d,{\bf {z}}_x}|\bf {y})}}\left[\log {q({\bf{y}}|{\bf z}_x,{\bf z}_d)} \right]\right],
   \end{aligned}
  \label{final_loss_funtion}
\end{equation}
where $\theta_y$ and $\varphi_y$ are the NS-VAE's encoder and decoder parameters, respectively. {In (\ref{final_loss_funtion}), KL divergence constraints for speech and noise latent variables are present. These constraints enable us to estimate the desired posterior distributions (${p({\bf z}_d|{\bf{y}})}$ and ${p({\bf z}_x|{\bf{y}})}$) from the noisy signal in a supervised manner. Furthermore, the inclusion of KL divergence terms ensures that the speech and noise signals can be effectively separated in the low-dimensional representation space.}

There are two stages for the PVAE-based SE algorithm. In the training stage, C-VAE and N-VAE are separately pre-trained by self-supervision using (\ref{clean_vae}) and (\ref{noise_vae}). After that, the C-VAE and N-VAE are frozen, and NS-VAE is trained by (\ref{final_loss_funtion}). In the enhancement stage, the NS-VAE encoder's two outputs can be used as the input of C-VAE and N-VAE to obtain the prior distributions $q({\bf{x}}|{\bf z}_x)$ and $q({\bf{d}}|{\bf z}_d)$ for SE.

\subsection{$\beta$-PVAE}
To further improve PVAE's SE performance, we propose to leverage $\beta$-VAE to improve PVAE's disentangling ability \cite{xiang2022deep} in our another preliminary work. Furthermore, the proposed $\beta$-PVAE makes the best use of the $\beta$-VAE's trade-off property to simplify the PVAE's network structure and training parameters by setting $\beta$ to infinity and discarding the noisy speech restoration term \cite{xiang2022deep}, which means that $\beta$-PVAE can achieve a better disentangling and enhancement performance than PVAE with a simpler structure. Based on our derivations \cite{xiang2022deep,xiang2022bayesian}, the $\beta$-PVAE's optimization target for  $\beta \to +\infty$ is \cite{xiang2022deep} 
\begin{equation}
 \begin{aligned}
  & \mathcal{L}_{\beta p} (\theta_y; {\bf y}) = {\mathbb E_{{\bf {y}} \sim p(\bf {y}),{\bf {x}} \sim p(\bf {x})}} \{D_{KL}\left({p({\bf z}_x|{\bf{y}})}||{p({\bf z}_x|{\bf{x}})}\right) \\
  & \quad \quad + {\mathbb E_{{{\bf {z}}_x} \sim p({{\bf {z}}_x}|{\bf {y}})}}[\log \frac{p({\bf z}_x|{\bf x})}{q({\bf z}_x)}]\} \\
  & \quad \quad + {\mathbb E_{{\bf {y}} \sim p({\bf {y}}), {\bf {d}} \sim p({\bf {d}})}} \{D_{KL}\left({p({\bf z}_d|{\bf{y}})}||{p({\bf z}_d|{\bf{d}})}\right) \\
  &  \quad \quad + {\mathbb E_{{{\bf {z}}_d} \sim p({{\bf {z}}_d}|{\bf {y}})}}[\log \frac{p({\bf z}_d|{\bf d})}{q({\bf z}_d)}]\}.
   \end{aligned}
  \label{beta_pvae}
\end{equation}
Comparing (\ref{beta_pvae}) with (\ref{final_loss_funtion}), we can find that there is no reconstruction term in $\beta$-PVAE. Thus, $\beta$-PVAE's framework can be simplified by removing the NS-decoder part (Fig.~\ref{fig:Bayesian_DNN}(b)). The $\beta$-PVAE's training process is similar to PVAE; the only difference is that the $\beta$-PVAE's training optimization target is~(\ref{beta_pvae}) rather than (\ref{final_loss_funtion}).


\subsection{Generative Adversarial Network (GAN)}

A GAN \cite{goodfellow2014generative} consists of two networks: a generator network and a discriminator network. The generator network $G(\mathbf{z})$ maps latent $\mathbf{z}$ ($\mathbf{z}\sim q(\mathbf{z})$) to the data space (e.g., observed signal data). Typically, there are no rigid restrictions for the distribution $q(\mathbf{z})$\cite{nowozin2016f}. The discriminator network $D(\cdot)$ is used to determine whether $\mathbf{y}$ is an actual training sample ($D(\mathbf{y})$) or it is generated by the model through $\mathbf{y}=G(\mathbf{z})$ ($D(G(\mathbf{z}))$). GANs can be optimized by different $f$-divergences \cite{nowozin2016f}. In Jensen–Shannon (JS) divergence, GANs is optimized by the minimax of the loss function \cite{goodfellow2014generative}:
\begin{equation}
 \begin{aligned}
  & \min_G\max_{D} \mathcal{L}_{gan}(G,D) = \\
  & \mathbb{E}_{{\mathbf{y}}\sim q_{data}({\mathbf{y}})}[\log(D({\mathbf{y}}))]+\mathbb{E}_{{\mathbf{z}}\sim q({\mathbf{z}})}[\log(1-D(G({\mathbf{z}})))].
  \label{GAN_JS_f}
   \end{aligned}
\end{equation}
GANs have been applied in SE \cite{pascual2017segan, michelsanti2017conditional,fu2021metricgan, ting2021speech}, but the researched methods do not consider how a good speech representation can be obtained as the input of the GAN for SE. Instead, they use the observed signal as the GAN's input to generate the speech signal \cite{pascual2017segan, michelsanti2017conditional}. Although there are no set restrictions for the GAN's input, an accurate and discriminative signal representation \cite{wang2018supervised} can usually lead to better generative performance for the GAN \cite{kim2021conditional, zhang2021visinger}.

\section{Speech Enhancement with VAE and GAN}

To obtain a higher quality enhanced speech, in this paper, we extend DRL-based SE framework \cite{xiang2022deep}. We propose a VAE-GAN SE algorithm which introduces adversarial training to increase the decoders' robustness and signal restoration ability. In this algorithm, we split the training process into two stages: the representation learning and the adversarial training. In the first, representation learning, stage, we leverage $\beta$-PVAE to disentangle speech and noise latent representations from the observed signal. The purpose is to obtain a good signal representation, making the clean speech generation easier.  In the second, adversarial training, stage, we freeze the $\beta$-PVAE's encoders and leverage adversarial training to optimize $\beta$-PVAE's decoders. GANs can generate desired samples without accurate knowledge of the input sample distribution \cite{goodfellow2014generative,nowozin2016f} (it only needs samples) and it can also improve VAE decoder's generative performance \cite{larsen2016autoencoding, parmar2021dual, huang2018introvae}, so GANs can mitigate the effect of potentially inaccurate posterior estimation for $\beta$-PVAE's decoders and improve decoder's generative ability. As a result, $\beta$-PVAE can achieve a satisfactory SE performance even if the posterior estimation is inaccurate. In this section, we will first show the details of representation learning. Then, we will explain the adversarial training processes. After that, we will indicate how to apply the proposed VAE-GAN to {conduct SE}.

\subsection{Stage 1: Representation Learning}

\begin{algorithm}[!tbp]
\caption{Representation Learning.}\label{alg:alg_rl}
\begin{algorithmic}
\STATE 
\STATE 
{\textbf{Pre-train 1:}} Using the speech dataset and loss function (\ref{clean_vae}) to train a general speech VAE (C-VAE) \cite{kingma2013auto}.  
\STATE 
{\textbf{Pre-train 2:}} Using the noise dataset and loss function (\ref{noise_vae}) to train a general noise VAE (N-VAE) \cite{kingma2013auto}.  
\STATE
\hspace{0.5cm}\textbf{Repeat:} 
\STATE 
\hspace{0.75cm} 1. Choose random $M$ samples from the speech, noise,
    \\ \hspace{0.75cm} and observed signal dataset and build a corresponding 
    \\ \hspace{0.75cm} mini-batch;
\STATE 
\hspace{0.75cm} 2. Use the chosen speech, noise, and observed signal 
    \\ \hspace{0.75cm} samples as the encoders' input of C-VAE, N-VAE, 
    \\ \hspace{0.75cm} and NS-VAE, respectively;
\STATE 
\hspace{0.75cm} 3. Estimate the related posterior probability  ${p({\bf z}_x|{\bf{y}})}$, \\ \hspace{0.75cm}  ${p({\bf z}_d|{\bf{y}})}, {p({\bf z}_x|{\bf{x}})}$, and ${p({\bf z}_d|{\bf{d}})}$ using the equations: \\ \hspace{0.80cm}
(1)\ \ ${\mu}_{\theta_{yx}}({\bf{y}}),{\sigma}_{\theta_{yx}}^2({\bf{y}}), {\mu}_{\theta_{yd}}({\bf{y}}), {\sigma}_{\theta_{yd}}^2({\bf y}) = G_{\theta_y}({\bf{y}})$, \\  \hspace{0.80cm}
(2)\ \ $ \small {\mu}_{\theta_{x}}({\bf{x}}), {\sigma}_{\theta_{x}}^{2}({\bf x}) = G_{\theta_x}({\bf{x}})$, \\  \hspace{0.80cm}
(3)\ \ $ \small {\mu}_{\theta_{d}}({\bf{d}}), {\sigma}_{\theta_{d}}^{2}({\bf d}) = G_{\theta_d}({\bf{d}})$; \\ \hspace{0.75cm}
4. Calculate loss function (\ref{beta_pvae}); \\ \hspace{0.75cm}
5. Freeze C-VAE and N-VAE and apply the SGD \\ \hspace{0.75cm} algorithm to update the NS-VAE's parameters $\theta_y$ \cite{kingma2013auto};
\STATE
\hspace{0.5cm}
\textbf{until the convergence of the loss function.}
\STATE 
\hspace{0.5cm}
\textbf{Return:} The trained NS-VAE ($G_{\theta_x}$).
\end{algorithmic}
\label{alg_rl}
\end{algorithm}

In the first stage, we aim to disentangle speech and noise latent variables from the observed signal. This process is accomplished by the proposed $\beta$-PVAE \cite{xiang2022deep}. {The purpose of the representation learning stage is to separate speech and noise signals in the low-dimensional representation space.}

In $\beta$-PVAE, C-VAE and N-VAE are optimized by (\ref{clean_vae}) and (\ref{noise_vae}), respectively, and NS-VAE is optimized by (\ref{beta_pvae}). To calculate (\ref{clean_vae}), (\ref{noise_vae}), and (\ref{beta_pvae}), it is necessary to determine the related posterior and prior distributions and predefine $q({{\bf z}_x})$ and $q({{\bf z}_d})$. For the simplicity of the calculation, we assume that all posterior and prior distributions are multivariate normal distributions with diagonal covariance~\cite{kingma2013auto}, which is similar to the previous VAE-based SE methods\cite{leglaive2018variance,bando2018statistical,leglaive2019semi,carbajal2021guided,fang2021variational}. For  NS-VAE, we have 
 \begin{equation}
   \begin{aligned}
   & {p({\bf z}_x|{\bf{y}})} = \mathcal{N} \left({{\bf {{\bf z}}}_x;{\mu}_{\theta_{yx}}({\bf{y}}),{\sigma}_{\theta_{yx}}^2({\bf y})\bf{I}}\right)\\
   & {p({\bf z}_d|{\bf{y}})} = \mathcal{N} \left({{\bf {{\bf z}}}_d;{\mu}_{\theta_{yd}}({\bf{y}}),{\sigma}_{\theta_{yd}}^2({\bf y})\bf{I}}\right),
  \end{aligned}
  \label{noisy_prior}
\end{equation}
where $\bf I$ is the identity matrix; ${\mu}_{\theta_{yx}}({\bf{y}}),{\sigma}_{\theta_{yx}}^2({\bf{y}}), {\mu}_{\theta_{yd}}({\bf{y}})$, and ${\sigma}_{\theta_{yd}}^2({\bf y})$ can be estimated by NS-VAE's encoder $G_{\theta_y}({\bf{y}})$ with parameter ${\theta_y}=\{\theta_{yx}, \theta_{yd}\} $. ${\mu}$ and ${\sigma}^2$ represent the mean and variance in the related Gaussian distributions, respectively. Moreover, the prior and posterior estimation for C-VAE is 
\begin{equation}
   \begin{aligned}
   & {p({\bf z}_x|{\bf{x}})} = \mathcal{N} \left({{\bf {{\bf z}}}_x;{\mu}_{\theta_{x}}({\bf{x}}),{\sigma}_{\theta_{x}}^2({\bf x})\bf{I}}\right) \\
   & q({\bf{x}}|{\bf z}_x) = \mathcal{N} \left({{\bf {{\bf x}}};{\mu}_{\varphi_{x}}({{\bf z}_x}),{\sigma}_{\varphi_{x}}^2({{\bf z}_x})\bf{I}}\right),
  \end{aligned}
  \label{posterior_c}
\end{equation}
where ${\mu}_{\theta_{x}}({\bf{x}})$ and ${\sigma}_{\theta_{x}}^2({\bf x})$ are obtained by C-VAE's encoder $G_{\theta_x}({\bf{x}})$ with parameter ${\theta_x}$, and ${\mu}_{\varphi_{x}}({{\bf z}_x})$ and ${\sigma}_{\varphi_{x}}^2({{\bf z}_x})$ can be estimated by C-VAE's decoder $G_{\varphi_x}({{\bf z}_x})$ with parameter ${\varphi_x}$. Similarly, for N-VAE, we have
\begin{equation}
   \begin{aligned}
   & {p({\bf z}_d|{\bf{d}})} = \mathcal{N} \left({{\bf {{\bf z}}}_d;{\mu}_{\theta_{d}}({\bf{d}}),{\sigma}_{\theta_{d}}^2({\bf d})\bf{I}}\right) \\
   & q({\bf{d}}|{\bf z}_d) = \mathcal{N} \left({{\bf {{\bf d}}};{\mu}_{\varphi_{d}}({{\bf z}_d}),{\sigma}_{\varphi_{d}}^2({{\bf z}_d})\bf{I}}\right),
  \end{aligned}
  \label{posterior_n}
\end{equation}
where ${\mu}_{\theta_{d}}({\bf{d}})$ and ${\sigma}_{\theta_{d}}^2({\bf d})$ are obtained by C-VAE's encoder $G_{\theta_d}({\bf{d}})$ with parameter ${\theta_x}$, and ${\mu}_{\varphi_{d}}({{\bf z}_d})$ and ${\sigma}_{\varphi_{d}}^2({{\bf z}_d})$ can be estimated by C-VAE's decoder $G_{\varphi_d}({{\bf z}_d})$ with parameter ${\varphi_d}$. Furthermore,  $q({{\bf {z}}_d})$ and $q({{\bf {z}}_x})$ are pre-defined as a centered isotropic multivariate Gaussian, which can be represented as 
\begin{equation}
   \begin{aligned}
   & q({\bf z}_x) = \mathcal{N} ({{\bf {{\bf z}}}_x;{\bf{0}},{\bf{I}}}) \\
   & q({\bf z}_d) = \mathcal{N} ({{\bf {{\bf z}}}_d;{\bf{0}},{\bf{I}}}).
  \end{aligned}
  \label{prior_single}
\end{equation}
The entire representation learning process is summarized in Algorithm~\ref{alg:alg_rl}.

\subsection{Stage 2: Adversarial Training}

\begin{figure*}[!tbp]
  \centering
  \centerline{\includegraphics[scale=0.8]{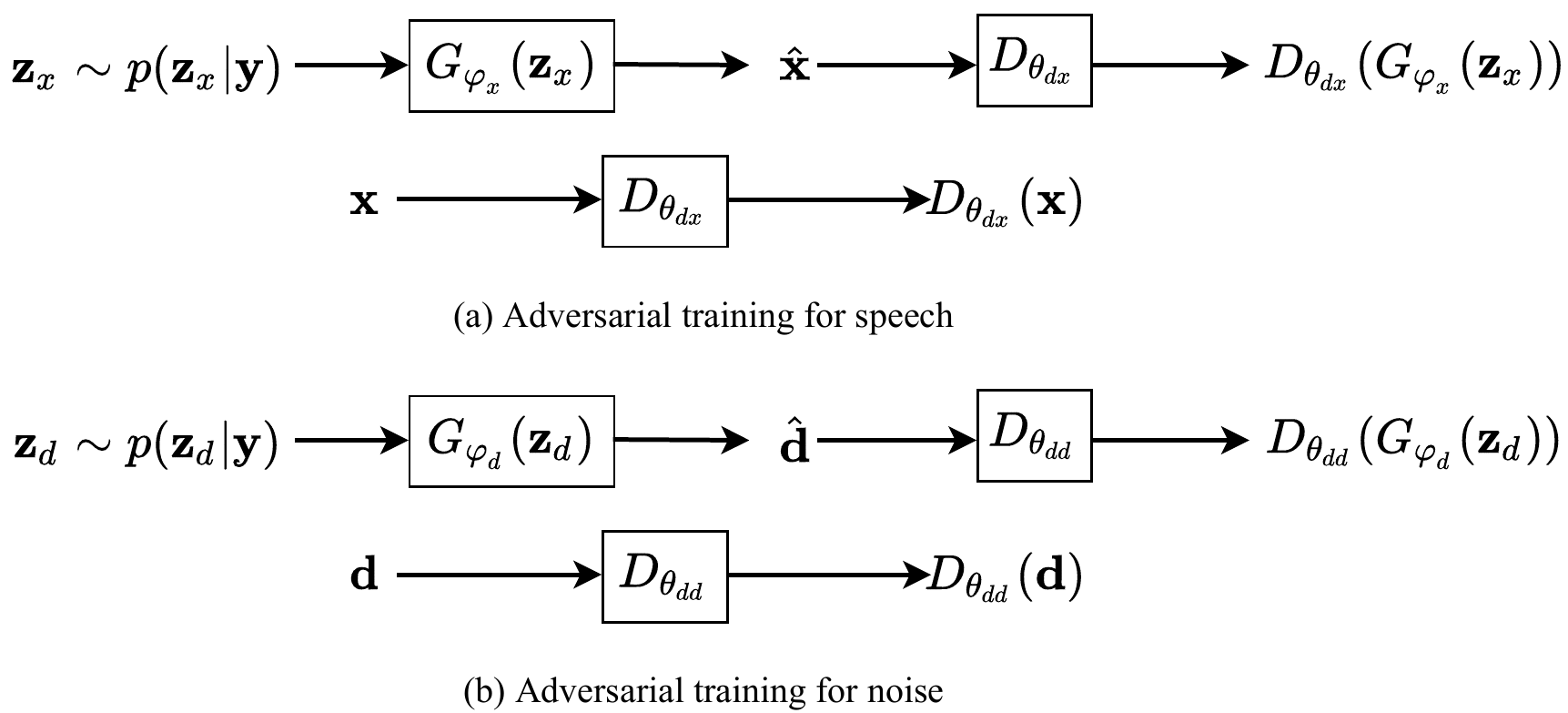}}
  \caption{Graphic illustration of adversarial training.}
  \label{fig:GAN}
\end{figure*}

\begin{algorithm}[!tbp]
\caption{Adversarial Training.}\label{alg:alg_ad}
\begin{algorithmic}
\STATE 
\hspace{0.5cm}\textbf{Repeat:} 
\STATE 
\hspace{0.75cm} 1. Choose random $M$ samples from the speech, \\ \hspace{0.75cm} noise, and observed signal dataset, respectively, and \\ \hspace{0.75cm} build a corresponding mini-batch;
\STATE 
\hspace{0.75cm} 2. Use the observed signal samples as the input of \\ \hspace{0.75cm} NS-VAE;
\STATE 
\hspace{0.75cm} 3. Estimate the related posterior probability  ${p({\bf z}_x|{\bf{y}})}$ \\ \hspace{0.75cm} and ${p({\bf z}_d|{\bf{y}})}$ using the following equation: \\ \hspace{0.80cm}
${\mu}_{\theta_{yx}}({\bf{y}}),{\sigma}_{\theta_{yx}}^2({\bf{y}}), {\mu}_{\theta_{yd}}({\bf{y}}), {\sigma}_{\theta_{yd}}^2({\bf y}) = G_{\theta_y}({\bf{y}})$;  \\ \hspace{0.75cm}
4. Apply the reparameterization trick to obtain sample \\ \hspace{0.75cm} ${{{\bf {z}}_x} \sim p({{\bf {z}}_x}|{\bf {y}})}$ and ${{{\bf {z}}_d} \sim p({{\bf {z}}_d}|{\bf {y}})}$  \cite{goodfellow2014generative};\\ \hspace{0.75cm} 
5. Use ${{{\bf {z}}_x}}$ and ${{{\bf {z}}_d}}$ as the C-VAE decoder's ($G_{\varphi_x}$) input \\ \hspace{0.75cm} and N-VAE decoder's ($G_{\varphi_d}$) input, respectively; \\ \hspace{0.75cm}
6. Calculate the loss function (\ref{GAN_gc}), (\ref{GAN_dc}), (\ref{GAN_gd}), (\ref{GAN_dd});  \\ \hspace{0.75cm}
5. Freeze all encoders and apply SGD to update\\ \hspace{0.75cm} parameters $\varphi_x$, $\varphi_d$,  $\theta_{dx}$, and $\theta_{dd}$  for $G_{\varphi_x}, G_{\varphi_d}$, $D_{\theta_{dx}}$, \\ \hspace{0.75cm} and $D_{\theta_{dd}}$ respectively; \\
\hspace{0.5cm}
\textbf{until the convergence of the loss function}
\STATE 
\hspace{0.5cm}
\textbf{Return:} The trained decoders and discriminators: \\ \hspace{0.5cm} $G_{\varphi_x}, G_{\varphi_d}$, $D_{\theta_{dx}}$, and $D_{\theta_{dd}}$.
\end{algorithmic}
\label{alg_ad}
\end{algorithm}

The second training stage aims to improve the decoders' robustness and signal restoration ability in $\beta$-PVAE for better SE performance. It is difficult to ensure that disentangled speech and noise latent representations are consistently accurate in complex noisy environments. Considering that decoders' SE performance relies on accurate representations, we propose to leverage adversarial training to mitigate this contradiction. In general, a GAN can generate the data, given the input is a random noise variable \cite{goodfellow2014generative, pascual2017segan}. Moreover, adversarial training can usually improve decoder's signal restoration ability ~\cite{larsen2016autoencoding, parmar2021dual, huang2018introvae}. As a result, we can use GANs to reduce decoders' dependence on accurate representation, which means that even with inaccurate representation estimations,  decoders can achieve a satisfactory SE performance. {Note that the signal separation process mainly occurs during the representation learning stage. In the adversarial training stage, the main role of the decoders is to convert low-dimensional representations back to high-dimensional signals, focusing on signal reconstruction rather than signal separation.}

To adopt adversarial training in the $\beta$-PVAE system, we add two discriminators, $D_{\theta_{dx}}(\cdot)$ and $D_{\theta_{dd}}(\cdot)$, with parameters $\theta_{dx}$ and $\theta_{dd}$, respectively. $D_{\theta_{dx}}(\cdot)$ is used to distinguish between the speech generated by the C-VAE decoder $G_{\varphi_x}({{\bf z}_x})$ and the ground truth speech $\bf{x}$. Similarly, we apply $D_{\theta_{dd}}(\cdot)$ to distinguish between the noise generated by the N-VAE decoder $G_{\varphi_d}({{\bf z}_d})$ and the ground truth noise $\bf{d}$. Fig.~\ref{fig:GAN} shows the related adversarial training process. In this work, we use the least squares GAN \cite{mao2017least} loss function for adversarial training, which has been widely used in various GAN applications \cite{kim2021conditional, zhang2021visinger} as it can achieve a more stable training process and avoid the problem of vanishing gradients,  compared to the original GAN \cite{goodfellow2014generative} loss function. Moreover, although GAN can generate high-quality signals, GAN may diverge too much from the target signals \cite{larsen2016autoencoding, parmar2021dual, huang2018introvae}. So, to ensure that the generated signals do not diverge too much from the ground truth signals, we reserve the original reconstruction term in the representation learning stage when conducting adversarial training. This is a GAN training trick for our proposed VAE-GAN, which is similar to the feature matching loss in previous applications of GANs \cite{kim2021conditional, zhang2021visinger, pascual2017segan, michelsanti2017conditional, kumar2019melgan, kong2020hifi}.  Therefore, the adversarial loss function for C-VAE-decoder can be expressed as follows: 
\begin{equation}
 \begin{aligned}
  & \mathcal{L}_{gan_c}(G_{\varphi_x}) = \mathbb{E}_{{{\bf {z}}_x} \sim p({{\bf {z}}_x}|{\bf {y}})}[(D_{\theta_{dx}}(G_{\varphi_x}({\mathbf{z}}_{x}))-1)^2] \\
  & \quad - {\mathbb E_{{\bf {z}}_x \sim p({{\bf {z}}_x}|{\bf{y}})}} [\log {q({\bf x}|{\bf z}_x)} ],
  \label{GAN_gc}
   \end{aligned}
\end{equation}
\begin{equation}
 \begin{aligned}
  & \mathcal{L}_{gan_c}(D_{\theta_{dx}}) = \mathbb{E}_{{{\bf {z}}_x} \sim p({{\bf {z}}_x}|{\bf {y}})}[(D_{\theta_{dx}}(G_{\varphi_x}({\mathbf{z}}_{x})))^2] \\
  & \quad + {\mathbb E_{{\bf {x}} \sim q_{data}({\bf{x}})}} [(D_{\theta_{dx}}({\bf{x}})-1)^2].
  \label{GAN_dc}
   \end{aligned}
\end{equation}
Similarly, the adversarial loss function for noise can be represented as 
\begin{equation}
 \begin{aligned}
  & \mathcal{L}_{gan_d}(G_{\varphi_d}) = \mathbb{E}_{{{\bf {z}}_d} \sim p({{\bf {z}}_d}|{\bf {y}})}[(D_{\theta_{dd}}(G_{\varphi_d}({\mathbf{z}}_{d}))-1)^2] \\
  & \quad - {\mathbb E_{{\bf {z}}_d \sim p({{\bf {z}}_d}|{\bf{y}})}} [\log {q({\bf d}|{\bf z}_d)} ],
  \label{GAN_gd}
   \end{aligned}
\end{equation}
\begin{equation}
 \begin{aligned}
  & \mathcal{L}_{gan_d}(D_{\theta_{dd}}) = \mathbb{E}_{{{\bf {z}}_d} \sim p({{\bf {z}}_d}|{\bf {y}})}[(D_{\theta_{dd}}(G_{\varphi_d}({\mathbf{z}}_{d})))^2] \\
  & \quad + {\mathbb E_{{\bf {d}} \sim q_{data}({\bf{d}})}} [(D_{\theta_{dd}}({\bf{d}})-1)^2].
  \label{GAN_dd}
   \end{aligned}
\end{equation}
The complete adversarial training process is summarized in
Algorithm 2.

\subsection{VAE-GAN for {Speech Enhancement}}

\begin{figure}[!tbp]
  \centering
  \centerline{\includegraphics[scale=0.60]{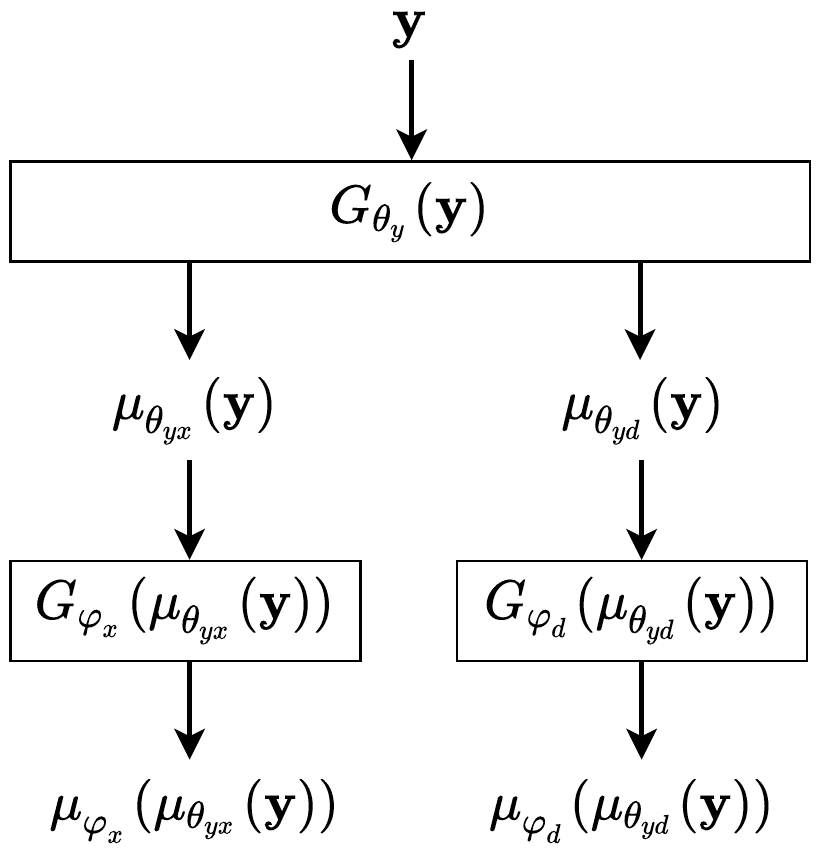}}
  \caption{VAE-GAN for online SE.}
  \label{fig:SE}
\end{figure}

\begin{algorithm}[!tbp]
\caption{{VAE-GAN-based SE.}}\label{alg:alg_se}
\begin{algorithmic}
\STATE 
\STATE 
1: Apply the observed signal ${\bf{y}}$ as the NS-VAE's encoder ($G_{\theta_x}$) input;
\STATE 
2. Estimate the posterior probability ${p({\bf z}_x|{\bf{y}})}$ and ${p({\bf z}_d|{\bf{y}})}$ by:
${\mu}_{\theta_{yx}}({\bf{y}}),{\sigma}_{\theta_{yx}}^2({\bf{y}}), {\mu}_{\theta_{yd}}({\bf{y}}), {\sigma}_{\theta_{yd}}^2({\bf y}) = G_{\theta_y}({\bf{y}})$;
\STATE
3. Use ${\mu}_{\theta_{yx}}({\bf{y}})$ and ${\mu}_{\theta_{yd}}({\bf{y}})$ as the inputs of C-VAE decoder $G_{\varphi_x}$ and N-VAE decoder $G_{\varphi_d}$, respectively;
\STATE
4. Apply decoders to estimate the speech and noise signal:
\STATE
\hspace{0.35cm} (1) ${\mu}_{\varphi_{x}}(\mu_{\theta_{yx}}({\bf{y}}))$, ${\sigma}_{\varphi_{x}}(\mu_{\theta_{yx}}({\bf{y}}))=G_{\varphi_x}(\mu_{\theta_{yx}}({\bf{y}}))$  
\STATE
\hspace{0.35cm} (2) ${\mu}_{\varphi_{d}}(\mu_{\theta_{yd}}({\bf{y}}))$, ${\sigma}_{\varphi_{d}}(\mu_{\theta_{yd}}({\bf{y}}))=G_{\varphi_d}(\mu_{\theta_{yd}}({\bf{y}}))$; 
\STATE
5. Use ${\mu}_{\varphi_{x}}(\mu_{\theta_{yx}}({\bf{y}}))$ and ${\mu}_{\varphi_{d}}(\mu_{\theta_{yd}}({\bf{y}}))$ as the estimated speech and noise signal;
\STATE
6. Apply waveform reconstruction \cite{xu2014regression} or mask the estimation \cite{wang2014training} to obtain the enhanced speech signal $\bf{\hat x}$.
\STATE
\textbf{Return:} The enhanced speech $\bf{\hat x}$.
\end{algorithmic}
\label{alg_se}
\end{algorithm}

The {SE stage} requires only the NS-VAE encoder $G_{\theta}$, C-VAE decoder $G_{\varphi_x}$, and N-VAE decoder $G_{\varphi_d}$ to conduct SE, which is similar to PVAE\cite{xiang2022bayesian} and $\beta$-PVAE \cite{xiang2022deep}. To obtain an enhanced signal, first, the observed signal is directly used as the input of $G_{\theta}$. Then, the posterior means ${\mu}_{\theta_{yx}}({\bf{y}})$ and ${\mu}_{\theta_{yd}}({\bf{y}})$ are obtained. After that, ${\mu}_{\theta_{yx}}({\bf{y}})$ and ${\mu}_{\theta_{yd}}({\bf{y}})$ are used separately as the input for $G_{\varphi_x}$ and $G_{\varphi_d}$ to estimate the speech mean ${\mu}_{\varphi_{x}}(\mu_{\theta_{yx}}({\bf{y}}))$ and noise mean ${\mu}_{\varphi_{d}}(\mu_{\theta_{yd}}({\bf{y}}))$, respectively. Finally, the estimated means are utilized as the enhanced speech and noise signal. The enhancement process is shown in Fig.~\ref{fig:SE} and Algorithm 3. In the {SE stage}, the means are used directly to estimate the signals, without the reparameterization trick \cite{kingma2013auto}, which is different from the training process \cite{kingma2013auto}. Moreover, the proposed VAE-GAN can simultaneously estimate the speech and noise in the observed signal, so the final enhanced signal can be obtained by direct waveform reconstruction \cite{xu2014regression} or mask estimation \cite{wang2014training}.

\section{Experimental Settings and Results}

In this section, the proposed VAE-GAN-based SE algorithm is evaluated. To explore VAE-GAN's SE potential, we use  related competitive algorithms as the reference methods to investigate VAE-GAN's SE performance.

\subsection{Datasets}

In this work, we created a training and test dataset using the speech and noise from the DNS challenge 2021 corpus~\cite{reddy2021interspeech}. To build a clean speech dataset, we selected English speakers and randomly split 70\% of the speakers for training, 20\% for validation, and 10\% for evaluation. For the noise, all the noise from the DNS noise corpus was randomly divided into training, validation, and test noise in a proportion similar to that used for speech utterances. The noise dataset comprised approximately 150 audio classes and 60,000 clips (the noise details can be found in \cite{reddy2021interspeech}). After that, the corresponding training, validation, and test corpus for speech and noise were randomly mixed using the DNS script \cite{reddy2021interspeech} with random signal-to-noise ratio (SNR) levels (between --10dB and 15dB). The other parameters of the signal mixing were the default values in the DNS script \cite{reddy2021interspeech}. Finally, we randomly chose 20 hours of mixed training utterances, 5 hours of mixed validation utterances, and 1 hour of mixed test utterances to build the experimental dataset. All signals were down-sampled to 16 kHz~\cite{reddy2021interspeech}.

We also used the LibriSpeech \cite{panayotov2015librispeech}, 100 environmental noises\cite{hu2010tandem}, and NOISEX-92 database\cite{varga1993assessment} to evaluate the SE performance of various algorithms. The purpose was to see the SE performance of various algorithms in the unseen dataset. Random one-hour speech data from LibriSpeech database were chosen and then mixed randomly with all noises from 100 environmental noises\cite{hu2010tandem} and the NOISEX-92 database\cite{varga1993assessment}. The mixed SNRs were randomly chosen from the --10dB to 15dB. Finally, we obtained a one-hour noisy speech test data.

\subsection{Experimental Setup}

\begin{figure}[!tbp]
  \centering
  \centerline{\includegraphics[scale=0.65]{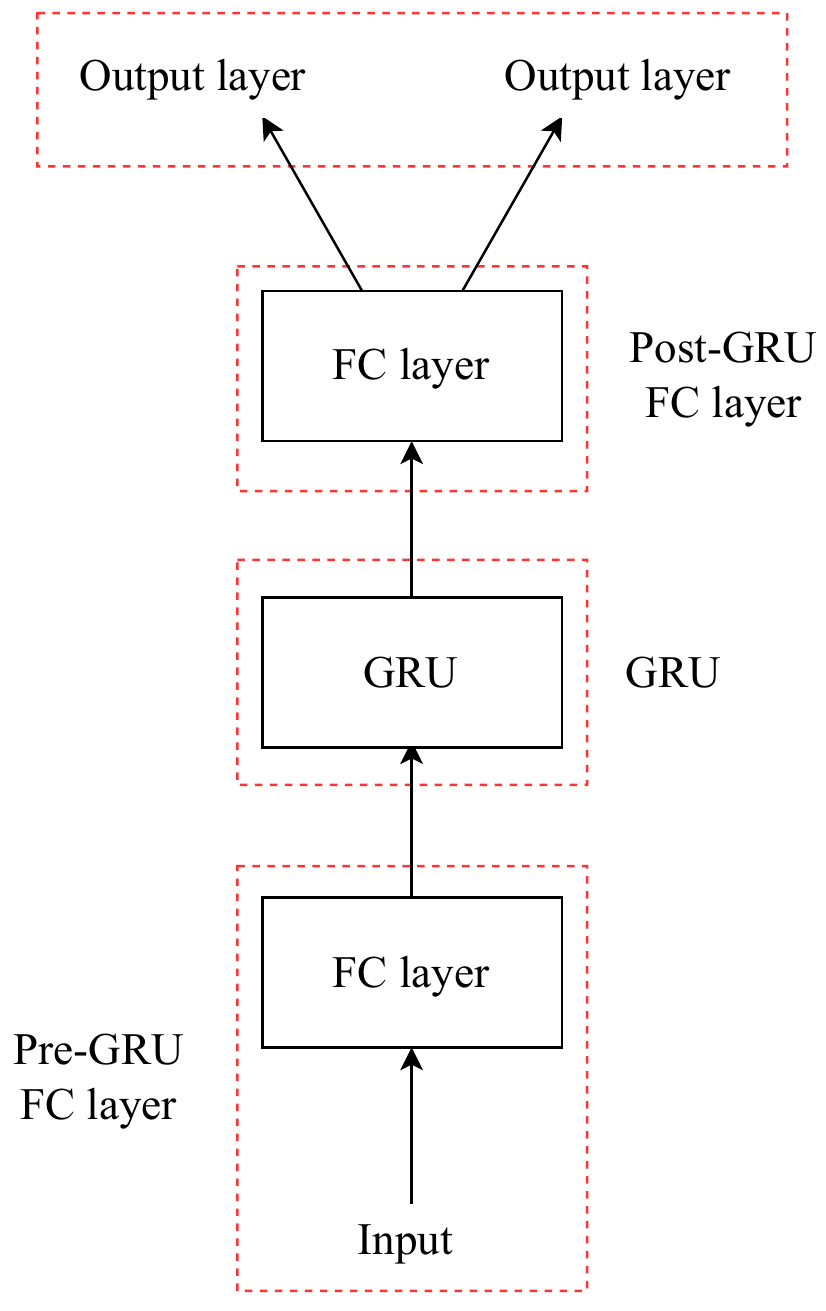}}
  \caption{Network structure in VAE-GAN.}
  \label{fig:GRU}
\end{figure}

\begin{table*}[!t]
 \centering
  \caption{Network Details of VAE-GAN}
  \label{tab: network_detail}
  \centering
 
  \begin{tabular}{ccccccccccccc}
    \toprule
    \multirow{2}*{Networks}& \multicolumn{3}{c}{Pre-GRU FC layer} &\multicolumn{2}{c}{GRU layer}&\multicolumn{3}{c}{Post-GRU FC layer}&\multicolumn{3}{c}{Output layer} \\

    \cmidrule(r){2-4} \cmidrule(r){5-6} \cmidrule(r){7-9} \cmidrule(r){10-12}  
    &Number &Nodes & AF & Number & Nodes & Number & Nodes & AF &Number & Nodes &AF\\
    \midrule
    $G_{\theta_x}$ and $G_{\theta_d}$ &3&257-512-512 & ReLU &1 & 512& 0 &N/A&N/A &2 &128&Linear \\
    \midrule
    $G_{\varphi_x}$ and $G_{\varphi_d}$ &1&128 & ReLU &1 & 512& 2 &512-512& ReLU & 2 & 257& Linear  \\
    \midrule
    $G_{\theta_y}$ & 3& 257-512-512 & ReLU & 1 & 512 & 1 & 512 & ReLU & 4 & 128 & Linear \\
    \midrule
    $D_{\theta_{dx}}$ and $D_{\theta_{dd}}$ & 2 & 257-512& ReLU & 1 & 256 & 1 & 512 & ReLU & 1 & 1 & Linear \\
    \bottomrule                             
  \end{tabular}
\end{table*}

In our experiment, the signal frame length was 512 samples (32 ms) with a frame shift of 256 samples. {A  STFT analysis was used to compute the DFT of each overlapping windowed frame.} The size of STFT was 256 points, so the 257-dimension LPS feature vectors were used to train the networks. Moreover, there were a total of 7 DNNs to be trained in VAE-GAN: C-VAE encoder $G_{\theta_x}$, C-VAE decoder $G_{\varphi_x}$, N-VAE encoder $G_{\theta_d}$, N-VAE decoder $G_{\varphi_d}$, NS-VAE encoder $G_{\theta_y}$, speech discriminators $D_{\theta_{dx}}$, and noise discriminator $D_{\theta_{dd}}$. All the DNNs in our experiment were based on the gated recurrent unit (GRU) \cite{cho2014learning} due to their computational efficiency and superior performance in SE \cite{reddy2019scalable}. In this work, we stacked GRU layers after the fully-connected (FC) layers, followed by hidden FC layers and FC output layers (Figure~\ref{fig:GRU}). This network design was similar to the baseline algorithm \cite{braun2020data} in DNS challenge 2022 \cite{dubey2022icassp}. The detailed model design of each neural network is shown in Table~\ref{tab: network_detail}, where AF represents the activation function in each output layer; Pre-GRU FC layer and Post-GRU FC layer represent the FC layer before the GRU layer and after the GRU layer, respectively; and the Nodes is the node number in each layer (all output layers have the same number of nodes in the same network). Additionally, we set the dimension of latent variables $L=128$, so for all encoders, the node number of the output layer is 128. All networks were trained by the Adam algorithm \cite{kingma2014adam} with a 128 mini-batch size. The learning rate is 0.001. {We conducted the experiments using the Python programming language and the PyTorch toolkit \cite{paszke2019pytorch}.}              

\subsection{Evaluation Metrics and Reference Methods }

\begin{table*}[!t]

 \centering
 
  \caption{{SI-SDR Comparison in DNS dataset with a 95\% confidence interval}}
  \label{tab: SI-SDR}
  \centering
    \begin{tabular}{cccccccccc}
    \toprule
    {SNR (dB)} & {Noise} & GAN-SE& {Y-SE-L} & {Y-SE-M} & NSNet2 & $\beta$-PVAE-L &VAE-GAN-L & $\beta$-PVAE-M &VAE-GAN-M \\
    & & \cite{michelsanti2017conditional}&\cite{huang2014deep} & \cite{huang2014deep} &\cite{braun2020data} & \cite{xiang2022deep} &  &\cite{xiang2022deep}&\\
    \midrule
    -5&-4.40&2.15& 1.88& 2.61 & 5.07&2.63&4.52&3.52&{\bf{5.37}}\\
      &($\pm\,$0.80)&($\pm\,$0.79)& ($\pm\,$0.78) & ($\pm\,$0.84) &($\pm\,$0.74)&($\pm\,$0.80)&($\pm\,$0.72)&($\pm\,$0.93)&($\pm\,${\bf{0.89}})\\
     \midrule
    0&2.63 &6.79 & 5.24 & 5.66 &9.77 &5.69 &8.48 &8.92 &{\bf{10.17}}\\
      &($\pm\,$1.04)&($\pm\,$0.61)& ($\pm\,$0.60) & ($\pm\,$0.89) &($\pm\,$0.81)&($\pm\,$0.59)&($\pm\,$0.52)&($\pm\,0.92$)&($\pm\,${\bf{0.86}})\\
     \midrule
    5&7.63 &9.30 & 7.02& 7.99&13.09 &8.10 &10.96 &12.96 &{\bf{ 14.11}}\\
      &($\pm\,1.08$)&($\pm\,$0.50)&($\pm\,$0.54) & ($\pm\,$0.88) &($\pm\,$0.82)&($\pm\,$0.46)&($\pm\,$0.39)&($\pm\,$0.93)&($\pm\,${\bf{0.85}})\\
     \midrule
    10&13.58 &11.75 & 9.02 & 10.16&16.76 &10.46 &13.07 &17.75 &{\bf{18.58}}\\
      &($\pm\,$1.05)&($\pm\,$0.42)& ($\pm\,$0.44) & ($\pm\,$0.81) &($\pm\,$0.72)&($\pm\,$0.35)&($\pm\,$0.30)&($\pm\,$0.88)&($\pm\,${\bf{0.84}})\\
    \midrule
    Average&4.86 &7.49& 5.79 &  6.61 &11.17 &6.72 &9.26 &10.78 &{\bf{ 12.06}}\\
      &($\pm\,$0.99)&($\pm\,$0.58)& ($\pm\,$0.59) & ($\pm\,$0.86) &($\pm\,$0.77)&($\pm\,$0.55)&($\pm\,$0.48)&($\pm\,$0.91)&($\pm\,${\bf{0.86}})\\
    
    \bottomrule                             
  \end{tabular}
  
\end{table*}

\begin{table*}[!t]
 \centering
 
  \caption{{STOI (\%) Comparison in DNS dataset with a 95\% confidence interval}}
  \label{tab: STOI}
  \centering
    \begin{tabular}{cccccccccc}
    \toprule
    {SNR (dB)} & {Noise} & GAN-SE& {Y-SE-L} & {Y-SE-M} & NSNet2 & $\beta$-PVAE-L &VAE-GAN-L & $\beta$-PVAE-M &VAE-GAN-M \\
    & & \cite{michelsanti2017conditional}&\cite{huang2014deep} & \cite{huang2014deep} &\cite{braun2020data} & \cite{xiang2022deep} &  &\cite{xiang2022deep}&\\
    \midrule
    -5&73.80 & 72.26& 71.13 & 72.44 &78.15 &72.94 &76.83 &77.27 &{\bf{79.29}}\\
      &($\pm\,$1.70)&($\pm\,$1.91)& ($\pm\,$1.89) & ($\pm\,$1.92) & ($\pm\,$1.61)&($\pm\,$1.77)&($\pm\,$1.81)&($\pm\,1.71$)&($\pm\,${\bf{1.80}})\\
     \midrule
    0&82.46 &81.47  & 81.02& 82.01 &87.03 &82.23 &85.62 &86.02 &{\bf{87.06}}\\
      &($\pm\,$1.40)&($\pm\,$1.42)& ($\pm\,$1.44) & ($\pm\,$1.51) & ($\pm\,$1.12)&($\pm\,$1.32)&($\pm\,$1.18)&($\pm\,$1.25)&($\pm\,${\bf{1.19}})\\
     \midrule
    5&88.01 &87.02  & 86.99 & 87.26 &91.63 &87.57 &90.71 &91.08 &{\bf{92.01}}\\
      &($\pm\,$1.11)&($\pm\,$1.02) & ($\pm\,$1.01) & ($\pm\,$0.93) &($\pm\,$0.81)&($\pm\,$0.99)&($\pm\,$0.80)&($\pm\,$0.91)&($\pm\,${\bf{0.82}})\\
     \midrule
    10&93.54 &92.13 & 92.01 & 92.92 &95.59 &92.54 &94.68 &95.58 &{\bf{96.02}}\\
      &($\pm\,$0.72)&($\pm\,$0.61) & ($\pm\,$0.71) &($\pm\,$0.73) &($\pm\,$0.47)&($\pm\,$0.59)&($\pm\,$0.46)&($\pm\,$0.51)&($\pm\,${\bf{0.47}})\\
     \midrule
    Average&84.45 & 83.22& 82.79 & 83.66 & 88.10 & 83.82 &86.96 &87.48 &{\bf{88.60}}\\
      &($\pm\,1.23$)&($\pm\,$1.24) & ($\pm\,$1.26) & ($\pm\,$1.27) &($\pm\,$1.09)&($\pm\,$1.00)&($\pm\,$1.06)&($\pm\,$1.09)&($\pm\,${\bf{1.07}})\\
      
    \bottomrule                             
  \end{tabular}
  
\end{table*}

     
      
  

\begin{table*}[!t]
 \centering
  \caption{{PESQ Comparison in DNS dataset with a 95\% confidence interval}}
  \label{tab: PESQ}
  \centering
    \begin{tabular}{cccccccccc}
    \toprule
    {SNR (dB)} & {Noise} & GAN-SE& {Y-SE-L} & {Y-SE-M} & NSNet2 & $\beta$-PVAE-L &VAE-GAN-L & $\beta$-PVAE-M &VAE-GAN-M \\
    & & \cite{michelsanti2017conditional}&\cite{huang2014deep} & \cite{huang2014deep} &\cite{braun2020data} & \cite{xiang2022deep} &  &\cite{xiang2022deep}&\\
    \midrule
    -5&1.81 &2.00 & 1.92 & 2.04 &2.28 &2.08 &{\bf{2.31}} &2.19 &2.30\\
      &($\pm\,$0.02)&($\pm\,$0.03)&($\pm\,$0.02) & ($\pm\,$0.03)& ($\pm\,$0.02)&($\pm\,$0.03)&($\pm\,$\bf{0.02})&($\pm\,$0.03)&($\pm\,${0.02})\\
     \midrule
    0&2.04 &2.33 & 2.31 & 2.40 &2.60 &2.46 &{\bf{2.64}} &2.55 &2.62\\
      &($\pm\,$0.02)&($\pm\,$0.02)&($\pm\,$0.03) & ($\pm\,$0.02) & ($\pm\,$0.02)&($\pm\,$0.03)&($\pm\,${\bf{0.02}})&($\pm\,$0.02)&($\pm\,$0.01)\\
     \midrule
    5& 2.28&2.62 & 2.61 & 2.70  &2.87 &2.77 &{\bf{2.94}} &2.85 &2.93\\
      &($\pm\,$0.02)&($\pm\,$0.02) &($\pm\,$0.02) &($\pm\,$0.02) &($\pm\,$0.02)&($\pm\,$0.02)&($\pm\,${\bf{0.01}})&($\pm\,$0.02)&($\pm\,$0.01)\\
     \midrule
    10&2.70 &3.00  & 3.01 & 3.12 &3.24 &3.14 &{\bf{3.29}} &3.21 &{\bf{3.29}}\\
      &($\pm\,$0.01)&($\pm\,$0.01) &($\pm\,$0.02) &($\pm\,$0.01) &($\pm\,$0.01)&($\pm\,$0.01)&($\pm\,${\bf{0.01}})&($\pm\,$0.01)&($\pm\,${\bf{0.01}})\\
      \midrule
    Average&2.21 &2.49 & 2.46 & 2.57 &2.75 &2.61 &{\bf{2.80}} &2.70 &2.79\\
      &($\pm\,$0.02)&($\pm\,$0.02) & ($\pm\,$0.02) & ($\pm\,$0.02) &($\pm\,$0.02)&($\pm\,$0.03)&($\pm\,${\bf{0.02}})&($\pm\,$0.02)&($\pm\,$0.01)\\
     
    \bottomrule
      
  \end{tabular}
  
\end{table*}

In this work, we will use the scale-invariant signal-to-distortion ratio (SI-SDR) in decibel (dB) \cite{le2019sdr}, short-time objective intelligibility (STOI)\cite{taal2011algorithm}, and perceptual evaluation of speech quality (PESQ)\cite{rix2001perceptual} as evaluation metrics to evaluate the proposed VAE-GAN's SE performance. SI-SDR is used to measure the signal distortion of the enhanced speech, so it can directly show the difference between the ground truth signal and the enhanced signal. PESQ and STOI are used to evaluate the quality and intelligibility for the enhanced speech, respectively. {To enhance the evaluation of speech enhancement (SE) performance on unseen datasets, we also employ DNSMOS P.835 \cite{reddy2021dnsmos,reddy2022dnsmos, naderi2021crowdsourcing}. This metric allows us to assess the speech quality (SIG), background noise quality (BAK), and overall quality (P808 MOS) of the audio samples. DNSMOS P.835 has been shown to highly align with human ratings for speech quality evaluation, making it a effective measure for our purposes.}

\begin{table*}[!t]
 \centering
 
  \caption{{ Experimental result comparisons in LibriSpeech dataset with a 95\% confidence interval}}
  \label{tab: LibriSpeech}
  \centering
    \begin{tabular}{cccccccccc}
    \toprule
    {Evaluation Metrics} & {Noise} & GAN-SE &Y-SE-L & Y-SE-M& NSNet2 & $\beta$-PVAE-L &VAE-GAN-L & $\beta$-PVAE-M &VAE-GAN-M \\
    & & \cite{michelsanti2017conditional}&\cite{huang2014deep} &\cite{huang2014deep} & \cite{braun2020data} & \cite{xiang2022deep} &  &\cite{xiang2022deep}&\\
    \midrule
    SI-SDR &1.81 &6.16 & 5.94 & 6.20 &9.20 &6.40 &8.24 &7.04 &{\bf{10.18}}\\
      &($\pm\,$0.23)&($\pm\,$0.36) & ($\pm\,$0.46) & ($\pm\,$0.52)& ($\pm\,$0.70)&($\pm\,$0.45)&($\pm\,$0.50)&($\pm\,0.46$)&($\pm\,${\bf{0.56}})\\
     \midrule
    STOI (\%) &82.75 &80.86 & 80.04 & 80.92  &86.03 &81.56 &84.50 &85.32 &{\bf{86.05}}\\
      &($\pm\,$1.63)&($\pm\,$1.69) &($\pm\,$1.71) & ($\pm\,$1.60)  &($\pm\,$1.51)&($\pm\,$1.53)&($\pm\,$1.47)&($\pm\,$1.53)&($\pm\,${\bf{1.50}})\\
     \midrule
    PESQ &2.31 &2.52  & 2.49 & 2.53 &2.69 &2.54 &2.71 &2.67 &{\bf{2.72}}\\
      &($\pm\,$0.03)&($\pm\,$0.02) & ($\pm\,$0.02) & ($\pm\,$0.03) &($\pm\,$0.01)&($\pm\,$0.03)&($\pm\,$0.02)&($\pm\,$0.02)&($\pm\,${\bf{0.01}})\\
       \midrule
    {SIG} & 2.87& 2.78  & 3.00  & 3.03  & 3.05  & 3.12 & 3.14 & 3.13  &{\bf{3.16 }}\\
      &($\pm\,0.05$)&($\pm\,$0.04) & ($\pm\,$0.04) & ($\pm\,$0.05) &($\pm\,$0.03)&($\pm\,$0.04)&($\pm\,$0.05)&($\pm\,$0.04)&($\pm\,${\bf{0.04}})\\
       \midrule
    {BAK} & 2.30 & 3.36  & 3.30  & 3.45  & 3.68 & 3.70 & 3.77  & 3.44  &{\bf{3.83 }}\\
      &($\pm\,$0.03)&($\pm\,$0.04) & ($\pm\,$0.03) & ($\pm\,$0.03) &($\pm\,$0.04)&($\pm\,$0.04)&($\pm\,$ 0.03)&($\pm\,$0.04)&($\pm\,${\bf{0.03}})\\
       \midrule
    {P808 MOS} & 2.92 &  3.11 &  3.08 &  3.16 & 3.44  & 3.18 & 3.49  & 3.30  &{\bf{ 3.62}}\\
      &($\pm\,$0.03)&($\pm\,$0.04) & ($\pm\,$0.05) & ($\pm\,$0.03) &($\pm\,$0.04)&($\pm\,$0.04)&($\pm\,$  0.03)&($\pm\,$0.04)&($\pm\,${\bf{0.04}})\\

    \bottomrule                             
  \end{tabular}
  
\end{table*}

To better evaluate the proposed VAE-GAN's SE performance, we choose three related competitive SE algorithms as reference methods. The first reference method is GAN-SE~\cite{michelsanti2017conditional}, which is a competitive GAN-based SE algorithm that can help us verify whether the better signal representations (disentangled and low-dimension representations) in the observed signal can improve GAN's SE performance. In addition, we can see the effectiveness of a disentangled signal representation for the GAN-based SE method. This also shows the DRL's importance for the DNN-based SE algorithm. The second reference method is $\beta$-PVAE~\cite{xiang2022deep}. By comparing VAE-GAN's SE performance with $\beta$-PVAE, we can validate our hypothesis that adversarial training can improve $\beta$-PVAE's SE performance (the $\beta$-PVAE's encoder and decoders have the same structure as the VAE-GAN). {In addition to the aforementioned methods, we also conducted a direct comparison between our proposed method and Y-SE \cite{huang2014deep}. Y-SE utilizes the same DNN-based SE architecture as our approach but is trained without the use of VAE and GAN. The only difference between Y-SE and our method lies in the training strategy. Y-SE is essentially an end-to-end trained model without the inclusion of specific disentanglement or DRL techniques. By comparing our method directly with Y-SE \cite{huang2014deep}, we can clearly observe the impact and benefits that our proposed approach brings to a general DNN-based SE framework.} Finally, we compare the proposed VAE-GAN with the DNS 2021 challenge baseline NSNet2 \cite{braun2020data, xia2020weighted} to see whether the VAE-GAN's SE performance is competitive with the current popular SE algorithms \cite{braun2020data}. The main purpose of our experiment is not to outperform all state-of-the-art (SOTA) performance, but to authentically verify the validity of the proposed VAE-GAN framework and its further potential.

{For the Y-SE, VAE-GAN and $\beta$-PVAE, enhanced speech can be obtained by waveform reconstruction \cite{xu2014regression} or mask estimation~\cite{wang2014training}.} The direct waveform reconstruction is based solely on the speech estimate, while the mask is based on both speech and noise estimate. So, we use $\beta$-PVAE-M and $\beta$-PVAE-L that represent that the enhanced speech is acquired by mask estimation and direct waveform reconstruction using $\beta$-PVAE \cite{xiang2022deep}, respectively; VAE-GAN-L and VAE-GAN-M denote that the enhanced speech is obtained by the proposed VAE-GAN using direct waveform reconstruction and mask estimation, respectively; { Y-SE-L and Y-SE-M denote that the enhanced speech is obtained by the Y-SE \cite{huang2014deep} using direct waveform reconstruction and mask estimation, respectively.}  We use the ideal ratio mask \cite{wang2014training} that is widely applied in various SE tasks \cite{wang2014training, sun2017multiple} to conduct mask estimation.

\subsection{Experimental Results and Analysis}

In this work, STOI, PESQ, and SI-SDR are used to evaluate the SE performance of SE algorithms. We show the experimental results at four representative SNR levels (-5dB, 0dB, 5dB, and 10dB): at each SNR level, we randomly select one hour of speech signal to conduct the evaluation.

Table~\ref{tab: SI-SDR} shows the SI-SDR comparison with a 95\% confidence interval in the DNS dataset \cite{reddy2021interspeech}. Comparing VAE-GAN-L and $\beta$-PVAE-L, it is evident that there is a SI-SDR score improvement, which illustrates that adversarial training can effectively improve the decoder's signal estimation performance and generate benefits for the signal reconstruction. Additionally, the performance of mask estimation depends on the accuracy of the signal estimation, so VAE-GAN-M also obtain higher SI-SDR score than $\beta$-PVAE-M.{ Comparing the VAE-GAN-based methods (VAE-GAN-L and VAE-GAN-M) with GAN-SE and Y-SE-based methods (Y-SE-L and Y-SE-M), we find that all VAE-GAN-based methods can achieve a higher SI-SDR score than GAN-SE and Y-SE-based methods, which indicates the importance of representation learning for some DNN-based SE frameworks. A disentangled signal representation can help GANs generate a 
higher quality target. This verifies our previous hypothesis.} Finally, considering that VAE-GAN-M also shows a higher SI-SDR score than NSNet2, the proposed algorithm is quite competitive with the current practical SE algorithms. In this paper, we choose only a basic DNN structure to conduct the related experiments. Based on the experimental results, we believe that our algorithm has a strong potential to achieve better SE performance if VAE-GAN is applied to a more advanced DNN structure \cite{hu20g_interspeech}.

The STOI comparisons in the DNS dataset \cite{reddy2021interspeech} are shown in Table~\ref{tab: STOI}, showing that VAE-GAN-based methods can continuously improve speech intelligibility from --5dB to 10dB. This finding is different from the $\beta$-PVAE-based method, in which it is difficult to improve the STOI score in high SNR environments. The comparison between VAE-GAN and $\beta$-PVAE indicates that adversarial training can effectively improve speech intelligibility. { Meanwhile, comparing VAE-GAN, GAN-SE and Y-SE-based methods, we find that VAE-GAN significantly outperforms other two methods, which demonstrates the importance of a good disentangled signal representation for improving speech intelligibility.} Additionally, Table~\ref{tab: STOI} indicates that VAE-GAN-M can also obtain higher STOI score than NSNet2.

Table~\ref{tab: PESQ} indicates the PESQ comparison with a 95\% confidence interval in the DNS dataset \cite{reddy2021interspeech}. Moreover, VAE-GAN-L can consistently obtain the highest PESQ score under all four SNR environments. Comparing VAE-GAN-L and $\beta$-PVAE-L, we find that VAE-GAN-L obtains a very significant PESQ score improvement (a 0.19 advantage for the average PESQ score.) by introducing adversarial training, which shows the importance of adversarial training in direct signal reconstruction that can mitigate the effects of inaccurate posterior estimation for signal estimation. In addition, it is of interest that VAE-GAN-L is competitive with VAE-GAN-M, a finding that is different from the previous SI-SDR and STOI comparisons. This may indicate that adversarial training is more suitable for improving speech quality \cite{michelsanti2017conditional}. Table~\ref{tab: PESQ} also shows that VAE-GAN-L achieves a higher average PESQ score than NSNet2  \cite{braun2020data} (a 0.05 advantage), which indicates the VAE-GAN's benefits for improving speech quality. {Finally, it is evident that representation learning is also very important for the GAN-based \cite{michelsanti2017conditional} and Y-SE-based SE algorithms \cite{huang2014deep}, improving speech quality (VAE-GAN-L outperforms GAN-SE with a 0.31 average PESQ score).} Here, we want to indicate that the PESQ results are very noteworthy because VAE-GAN-L-based method that is without noise and mask estimation can outperform the mask-based method VAE-GAN-M. {In general, the mask or filter-based methods \cite{jensen2015noise,christensen2016experimental} need to estimate the speech and noise signal, or directly predict masks or complex filters for SE.} However, based on the experimental results, maybe we need to consider whether we still need to apply mask or filter for SE if we can use DRL or other methods to estimate high-quality speech signals because the filter or mask may also damage the speech signal \cite{jensen2015noise}. This problem will be considered in our following research. 

{In conclusion, by comparing our VAE-GAN-based method with the Y-SE-based method, we can clearly observe the significant impact and benefits that our proposed approach brings to the general DNN-based speech enhancement (SE) framework. This comparison effectively demonstrates the added value of incorporating VAE and GAN in the training process. The use of VAE helps in learning meaningful representations and disentangling latent variables, while GAN enhances the robustness and generative capabilities of the model. Together, these components contribute to the superior performance and improved results achieved by our proposed approach.}

Table~\ref{tab: LibriSpeech} presents the experimental comparisons in the LibriSpeech dataset \cite{panayotov2015librispeech} featuring the average scores of different SNR levels. The results in the LibriSpeech dataset tend to be similar to the results in the DNS dataset \cite{reddy2021interspeech}, which indicates that the proposed algorithm can still achieve satisfactory SE performance for unseen signals. Comparing $\beta$-PVAE-L and VAE-GAN-L, it is evident that VAE-GAN-L  returns higher SI-SDR, STOI, and PESQ scores than $\beta$-PVAE-L, supporting the importance of adversarial training for improving the accuracy of signal estimation.  Furthermore, as previously, VAE-GAN-M can produce the best SE performance. {Moreover, when comparing VAE-GAN-M with NSNet2 using DNSMOS P.835 evaluation metrics (SIG, BAK, and P808 MOS), we observe that VAE-GAN-M exhibits a notable advantage in enhancing the human listening experience. These results show the benefits of our algorithm in improving the subjective perception of speech quality.}

To sum up, we find that the proposed VAE-GAN can achieve the best SE performance compared with the reference methods under the STOI, PESQ, and SI-SDR evaluation metrics. The experimental results demonstrate that: 1) representation learning can help the GAN-based SE method to obtain better SE performance; 2) adversarial training can significantly improve decoders' signal estimation in $\beta$-PVAE. Moreover, the comparison of VAE-GAN and NSNet2 \cite{braun2020data} shows that VAE-GAN is very competitive with the current SE algorithms \cite{braun2020data, xia2020weighted}. In this experiment, we only use a basic neural network structure \cite{braun2020data}; however, based on the experimental results, we believe that VAE-GAN has a significant potential to achieve better SE performance provided VAE-GAN is applied in more advanced neural networks \cite{trabelsi1705deep,choi2018phase,ronneberger2015u}.

\section{Conclusion and Future Work}

In this paper, we propose a two-stage DRL-based (VAE-GAN) SE algorithm. VAE-GAN leverages adversarial training to mitigate the problem of inaccurate posterior estimation in $\beta$-PVAE and can reduce the effect of inaccurate posterior estimation towards signal reconstruction, resulting in a more accurate speech estimation from the observed signal. We also compare the proposed VAE-GAN with other related competitive SE algorithms, and the experimental results show that VAE-GAN can obtain higher STOI, PESQ, and SI-SDR scores and achieve the best SE performance among the competing algorithms. Therefore, the results verify that DRL can significantly improve SE performance for the GAN-based SE method \cite{michelsanti2017conditional}, which validates DRL's importance for SE. On the other hand, the results also support that adversarial training is crucial for improving $\beta$-PVAE's SE performance. According to the experiments, VAE-GAN can have a significant potential in achieving better SE performance if applied in other advanced neural network structures. 

For future work, we propose two ways which may further improve VAE-GAN's SE performance. First, as mentioned before, it is possible to apply the proposed VAE-GAN in more advanced neural network structures. For example, we can consider using complex neural networks \cite{hu20g_interspeech,trabelsi1705deep,choi2018phase,ronneberger2015u} to perform related prior and posterior estimations in VAE-GAN with complex Gaussian distributions. {In addition, we can also apply real-world recordings to evaluate the SE performance of related SE algorithms.} Second, the proposed VAE-GAN can disentangle different types of latent variables, so it can possible to disentangle phoneme or emotional latent variables from the observed signal, which means it can be possible to analyze context information when conducting SE, a probability that has not been considered in previous SE methods \cite{loizou2013speech,wang2018supervised}. Finally, additional SE-related information can be considered to achieve better SE performance.

\section*{Acknowledgments}
This work was supported by Innovation Fund Denmark (Grant No. 9065-00046).


\bibliographystyle{IEEEtran}
\bibliography{IEEEabrv,myabrv_new,my_reference}

\end{document}